\documentclass[aps,prb, 10pt, twocolumn]{revtex4-2}
\usepackage{amsmath}
\usepackage{amsfonts}
\usepackage{amssymb}
\usepackage{graphicx}
\usepackage{bm}
\usepackage{hyperref}
\usepackage{xcolor}
\usepackage{booktabs}

\begin{document}
%%%%%%%%%%%%%%%%%%%%%%%%%%%%%%%%%%%%%%%%%%%%%%%%%%%%%%%%%%%%%%%%

%\title{Quantum Gaussian Processes based Machine Learning: A Proof-of-Concept and Use-Case for Line Parameter Estimation in Electrical Grids}
\title{Quantum multi-output Gaussian Processes based Machine Learning for Line Parameter Estimation in Electrical Grids}

\author{Priyanka Arkalgud Ganeshamurthy$^{1}$}
\email{priyanka.ag@eonerc.rwth-aachen.de}

\author{Kumar Ghosh$^{2}$}
\email{jb.ghosh@outlook.com}

\author{Corey O' Meara$^{2}$}
\email{corey.o'meara@eon.com}

\author{Giorgio Cortiana$^{2}$}

\author{Jan Schiefelbein-Lach$^{3}$}

\author{Antonello Monti$^{1}$}

\affiliation{$^1$~E.ON Energy Research Center, Institute for Automation of Complex Power Systems, RWTH Aachen University, Aachen, Germany \\
$^2$~E.ON Digital Technology GmbH, Hannover, Germany.\\
$^3$~E.ON Group Innovation GmbH, Essen, Germany. }

%\author{\uppercase{Priyanka Arkalgud Ganeshamurthy Kumar Ghosh \authorrefmark}, Corey O' Meara \authorrefmark{2}, Giorgio Cortiana \authorrefmark{2}, Jan Schiefelbein-Lach \authorrefmark{3}, Antonello Monti \authorrefmark{1} }
%}

%\address[1]{E.ON Energy Research Center, Institute for Automation of Complex Power Systems, RWTH Aachen University, Aachen, Germany (e-mail: priyanka.ag@eonerc.rwth-aachen.de)}
%\address[2]{E.ON Digital Technology GmbH, Hannover, Germany. (e-mail: kumar.ghosh@eon.com)}
%\address[3]{E.ON Group Innovation GmbH, Essen, Germany. (e-mail: jan.schiefelbein-lach@eon.com)}

%%%%%%%%%%%%%%%%%%%%%%%%%%%%%%%%%%%%%%%%%%%%%%%%%%%%%%%%%%%%%%%%%%%%%%%%%%%%%

\begin{abstract}
Gaussian process (GP) is a powerful modeling method with applications in machine learning for various engineering and non-engineering fields. Despite numerous benefits of modeling using GPs, the computational complexity associated with GPs demanding immense resources make their practical usage highly challenging. In this article, we develop a quantum version of multi-output Gaussian Process (QGP) by implementing a well-known quantum algorithm called HHL, to perform the Kernel matrix inversion within the Gaussian Process. To reduce the large circuit depth of HHL a circuit optimization technique called Approximate Quantum Compiling (AQC) has been implemented. We further showcase the application of QGP for a real-world problem to estimate line parameters of an electrical grid. Using AQC, up to 13-qubit HHL circuit has been implemented for a 32x32 kernel matrix inversion on IBM Quantum hardware for demonstrating QGP based line parameter estimation experimentally. Finally, we compare its performance against noise-less quantum simulators and classical computation results.
\end{abstract}

\maketitle

%%%%%%%%%%%%%%%%%%%%%%%%%%%%%%%%%%%%%%%%%%%%%%%%%%%%%%%%%%%%%%%%%%%%%%%%%%%%%

\section{Introduction}
\label{sec:introduction} 
% Use of machine learning for PS applications; use of GP for PS applications; 
In recent times, extracting useful information from the immense volume of data has been of prime significance in technological evolution. Gaussian processes (GPs) are a popular modeling technique, which provides a distribution of function along with mean and covariance of the function to be modeled. GP modeling are a popular choice for machine learning due to it's non-parametric data-driven nature, inherent quantification of uncertainty, kernel based flexibility allowing modeling data of different nature, allowing online belief updating due to Bayesian framework and interpretability \cite{GP_Rasmussen}. These numerous benefits showcase versatility of GP which have found applications in several fields ranging from finance, chemistry, battery technology, robotics to power systems, to name a few. For example, in the area of power systems, GP modeling have been used for estimating network line parameters in \cite{gp-lpe}, for inferring of synchronous generator dynamics in \cite{DD_SE_28}, for performing probabilistic load-flow allowing use of non-Gaussian measurement errors in \cite{GP_PF} and for various state estimation objectives in \cite{DD_SE_29}, \cite{DD_SE_19} and \cite{DD_SE_20}. 
Despite numerous merits, the data-driven approaches made possible by GPs face a major hurdle for their practical implementation in terms of limited scalability due to high computational complexity \cite{GP_Rasmussen}.    

In the last decade Quantum computing \cite{Nielsen} has been a center of attention due to its ability to outperform classical computation and information processing. Many known quantum algorithms have various applications, such as, search algorithm \cite{Grover}, integer factorization \cite{Shor}, solving constraint satisfaction problems \cite{Montanaro}, and quantum machine learning \cite{Mohseni, Biamonte, qsvm}. 
Recognising the drawback of GP and theoretical speed-up promised by certain quantum algorithms, authors of \cite{zhao2019quantum} suggested the use of HHL algorithm for implementing quantum algorithm for GP. 
In this article, we develop a quantum version of Gaussian process (QGP) where the Kernel matrix inversion during the training phase of GP is replaced by the HHL algorithm. Since HHL is well known for it's huge circuit depth, we implement another another quantum algorithm Approximate Quantum Compiling (AQC) to reduce the circuit depth of HHL to an extent such that it is implementable in the current quantum hardware. Further, we apply QGP for an engineering problem of estimating electrical grid line parameter and for prediction of measured electrical quantities by building on previous work in \cite{gp-lpe}, where multi-output GPs are trained simultaneously in a single formulation. To showcase a prototype of this QGP application, we present the results of execution on a real quantum hardware, named IBM Auckland, and compare our results with classical GP.

The main contributions of this article are the following:
\begin{itemize}
	\item Implementation of multi-output QGP, where the Kernel matrix inversion is replaced by HHL algorithm. To deal with high circuit depth, the HHL algorithm is complemented with AQC.
	\item During applying HHL the condition number of the kernel matrix appears to be very high, therefore we applied a matrix conditioning scheme; see subsection \ref{subsec:qgp_implementation} for more details about the above two points.
	\item Adaptation of QGP to a real-world problem of line parameter estimation and measurement signal predictions in electrical grids. 
	\item Demonstration of QGP based line parameter estimation and measurement signal predictions on a real quantum hardware IBM Auckland (NISQ).
	\item HHL algorithm is applied to invert 32x32 matrix using up to 13 qubits quantum circuit on real IBM quantum hardware. On similar lines, authors in \cite{qc_pf_real_test} have shown HHL on 2x2 matrix with 5 qubits on real quantum hardware.
\end{itemize}

The rest of the article is organized into five main sections. Section \ref{sec:gp_back_concept} introduces background concepts of GP. Section \ref{sec:qgp} presents the QGP algrithm. Section \ref{sec:gp-lpe} introduces the mathematical formulation and test-set up for line parameter estimation as well as the adaptation of QGP for the same. Section \ref{sec:results_diss} presents results and discussions on the comparison of line parameter estimation using classical and quantum GP variants. Finally Section \ref{sec:conclusion} concludes the article. 

%Gaussian processes (GPs) are a widely used model for regression problems machine learning. Implementation of GP regression typically requires $\mathcal{O}(n^3)$ logic gates. In article \cite{zhao2019quantum}, Zhao et al.  introduced a quantum inspired Gaussian process regression (GPR), leading to a reduction in computation time. Later Kus et al., in article \cite{kus2021sparse}, proposed a quantum circuit for implementing quantum Gaussian processes defined in \cite{kus2021sparse} and use quantum phase estimation to induce a low-rank approximation analogous to that in classical sparse Gaussian processes. Previously, Das et al. in article \cite{das2018continuous}, introduced an algorithm for Gaussian process regression using continuous-variable quantum systems that can be realized with technology based on photonic quantum computers under certain assumptions regarding distribution of data and availability of efficient quantum access.

\section{Background Concepts}
\label{sec:gp_back_concept}
\subsection{Gaussian Processes}
\label{sec:gp_explained}
\noindent 
A Gaussian Process is defined as a collection of indexed random variables, any finite number of which have (consistent) joint Gaussian distributions \cite{GP_Rasmussen}. %\cite{b0}. 
GP can be used for regression and classification applications as a supervised non-parametric Bayesian machine learning technique \cite{GP_Rasmussen}. In this section we briefly describe the technique of GP training and its use for regression. 

Consider a random process $\mathbf{f(\mathbf{t})}$ to be modeled as GP, which takes $D$ dimension input vector $\mathbf{t} = [t^1, t^2, ..., t^D]$. To learn the random process $\mathbf{f(\mathbf{t})}$, the training set involves $N$ samples of $D$ dimensional input $\mathbf{T}_{N \times D} = [\mathbf{t_1}, \mathbf{t_2},..., \mathbf{t_N}]^{'}$ and its corresponding realization of the function is denoted by $\mathbf{f(T)} = [f(\mathbf{t_1}), f(\mathbf{t_2}),..., f(\mathbf{t_N})]^{'}$.  
If we assume that the random process model output is a realization of a GP, then by definition, the finite collection of random variables given by $\mathbf{f(t)}$ will follow a joint multivariate normal probability distribution given by:  

\begin{align}
	\scriptsize
	\begin{bmatrix}
		\mathbf{f(t_1)}\\
		\mathbf{f(t_2)}\\
		.\\
		.\\
		\mathbf{f(t_N)}
	\end{bmatrix}
	~ \sim\mathcal{N}\left(
	\begin{bmatrix}
		\mathbf{m(\mathbf{t_1})}\\
		\mathbf{m(\mathbf{t_2})}\\
		.\\
		.\\
		\mathbf{m(\mathbf{t_N})}
	\end{bmatrix},
	\begin{bmatrix}
		k(\mathbf{t_1},\mathbf{t_2}) & ... &k(\mathbf{t_1},\mathbf{t_N})\\
		... &  \ddots & ...\\
		k(\mathbf{t_N},\mathbf{t_1}) &... &k(\mathbf{t_N},\mathbf{t_N})
	\end{bmatrix}
	\right) .
	\label{mvn}
\end{align}
This is a Gaussian distribution over vectors, given the number of samples are finite. However, when the number of samples $N$ is increased, it leads to a GP, which is a distribution over functions. A GP is therefore fully specified by a mean function and a covariance function and can be written as:

\begin{align*}
	%\scriptsize
	\mathbf{f(T)} ~& \sim\mathcal{GP}(\mathbf{m(\mathbf{T})}, \mathbf{K(T,T)})
\end{align*}
where, $\mathbf{m(\mathbf{T})}$ denotes mean function and $\mathbf{K(T,T)}$ denotes kernel function representing the covariance of the GP.

The observations of function output $\mathbf{f(t)}$ in reality are noisy. Here it is assumed that the observations of model $\mathbf{f(\mathbf{t})}$ contains Gaussian white noise as given by \eqref{eq2}. The observations of random process $\mathbf{f(t)}$ for $N$ samples is denoted by $\mathbf{y}_{N \times 1} = [y_1, y_2,..., y_N]$.

\begin{align}
	%\scriptsize
	\begin{split}
		\mathbf{y} &= \mathbf{f(\mathbf{t})} + \mathbf{\epsilon} ;   \:\:\:\:\:\:\:\:\:\:\:\:\:\:\:\:\:\:\:\:\:\:\:\:\:\:\:\:\:\:  \mathbf{\epsilon} ~ \sim\mathcal{N}(\mathbf{0}, {\sigma_n}^2\mathbf{I}) \\
		\mathbf{y} &~ \sim\mathcal{GP}(\mathbf{m(\mathbf{T})}, \mathbf{K(T,T)} + {\sigma_n}^2\mathbf{I}) 
		\label{eq2}
	\end{split}
\end{align}
where $\sigma_n^2$ is variance of the noise and  $\mathbf{I}$ is identity matrix.

The kernel function defines the covariance of the GP and dictates the relationship between the variation of input to the variation of output quantities of the random process. The selection of the kernel function depends upon prior understanding of the process to be modeled. One such popular kernel is the Radial Basis Functions (RBF) kernel, given by \eqref{rbf}. 
%and it provides smoothness and it gives rise to functions which are infinitely differentiable \cite{GP_Rasmussen, GP_01}. %\cite{b0}, \cite{b1}. 
%Each $k(\mathbf{.,.})$ term in \eqref{mvn} can be evaluated using \eqref{rbf}. 
RBF kernel parameter $\sigma^2$ is a variance and $w_d$ is the Automatic Relevance Determination weights which indicates characteristic length scale for each input dimensions $d$. 

\begin{align}
	\scriptsize
	k(t,t^{'}) = \sigma^2 exp\left(-0.5\sum_{d=0}^{D} w_d(t_d - {t_d}^{'})^2 \right) .
	\label{rbf}
\end{align}
These kernel parameters form the set of hyper-parameters $\phi = ({\sigma}^2, w_d)$, which are learnt with the help of training data. The optimal hyper-parameters $\phi$ are determined by maximizing the negative log marginal likelihood given by \ref{MLE}. This step completes the training of GP model. 

\begin{align}
	%\scriptsize
	\begin{split}
		-\log p(\mathbf{y}|\mathbf{T}, \phi, {\sigma_n}^2) &= \frac{1}{2} \mathbf{y}^{'}(\mathbf{K} + {\sigma_n}^2\mathbf{I})^{-1}\mathbf{y}   \\
		&  + \frac{1}{2} \log |\mathbf{K} + {\sigma_n}^2\mathbf{I}| + \frac{N}{2}\log 2\pi
		\label{MLE}
	\end{split}
\end{align}
where $\mathbf{K(T,T)}$ is simplified and written as $\mathbf{K}$. 
%To put it in simple terms, the model fitting problem in GP is to "extract" information about the model from the sample training data provided and mathematically "encode" it into kernel function parameters called hyper-parameters. This makes the GP learning method interpretable, since the hyper-parameters after training contain subtle information about the model learnt.
After the training, for a given new test inputs $\mathbf{T_*}$, we can predict its corresponding output $\mathbf{f(\mathbf{T_*})}$ by expressing  $ \mathbf{f(\mathbf{T})}$ and $ \mathbf{f(\mathbf{T_*})}$ as joint distribution given by:

\begin{align}
	%\scriptsize
	\begin{bmatrix}
		\mathbf{f(T)}\\
		\mathbf{f(T_*)}
	\end{bmatrix}
	~ \sim\mathcal{N}\left(
	\begin{bmatrix}
		\mathbf{m(\mathbf{T})}\\
		\mathbf{m(\mathbf{T_*})}
	\end{bmatrix},
	\begin{bmatrix}
		\mathbf{K} + {\sigma_n}^2\mathbf{I} & \mathbf{K_*}\\
		\mathbf{K_*^{'}} & \mathbf{K_{**}}
	\end{bmatrix}
	\right)
	\label{predic}
\end{align}
where, $ \mathbf{K} = \mathbf{K(T,T)}$, $  \mathbf{K_*} = \mathbf{K(T,T_*)}$, $  \mathbf{K_{**}} = \mathbf{K(T_*,T_*)}$. 
The conditional distribution provides the predictive mean and covariance of the output $ \mathbf{f(\mathbf{T_*})}$ for test input $ \mathbf{T_*}$ as follows:

\begin{align}
	%\scriptsize
	\begin{split}
		\mathbf{f(T_*)}|\mathbf{T,y,T_*} &~ \sim\mathcal{N} (\mathbf{m(T_*)}, cov(\mathbf{f(T_*)})) \\
		\mathbf{m(T_*)} &= \mathbf{K_*^{'}}[\mathbf{K} + {\sigma_n}^2\mathbf{I}]^{-1}\mathbf{y} \\
		cov(\mathbf{f(T_*)}) &= \mathbf{K_{**}} - \mathbf{K_*^{'}}[\mathbf{K} + {\sigma_n}^2\mathbf{I}]^{-1}\mathbf{K_*} .
		\label{predict}
	\end{split}
\end{align}

%\subsection{Linear Operations on Gaussian Processes}
%\label{sec:linear_op_gp}
GP method has several properties which makes them a popular modeling choice in machine-learning applications. One such advantage is that the GP models are closed under operations such as summation and linear operation. Linear operation $\mathcal{L}$ on a GP model, such as scaling, differentiation or integration also leads to a GP. 

\begin{align*}
	\mathcal{L}\mathbf{f(.)} &~  \sim\mathcal{GP}(\mathcal{L}\mathbf{m(\mathbf{.})}, \mathcal{L}^2\mathbf{K(.,.)}) .
\end{align*}

Recent advancements on Gaussian process techniques have shown that this special property of closed form on linear operations can be further exploited to estimate the parameters of a linear equation \cite{GP_01}. For this, the kernel matrix of the joint distribution of all variables in the linear equation is derived based on the form of the equation starting with a base kernle of a variable in the linear equation. This way, the parameters of the linear equations become part of the kernel structure, and therefore can be optimized using negative log-likelihood along with hyper-parameters of the base kernel. 

\iffalse
Other benefits and features of this method as seen from power system domain perspective are as follows:
\begin{enumerate}
	\item Method can work for measurements with different sampling rate
	\item The method also inherently offers capability of prediction of measured quantities
	\item By adjusting the problem formulation, for a single-phase two-terminal line, the method can work with 3 measured quantities instead of 4 quantities
	\item Correlation between measurements can be considered due to the physical equations 
	\item The method is completely decoupled from state estimation routine
\end{enumerate}
\fi

\subsection{Computational Complexity of Gaussian Proceses}
As described in Section \ref{sec:gp_explained}, the GP hyper parameters are learnt by optimizing the negative log marginal likelihood given by \ref{MLE}. This optimization can be solved  using a gradient optimizer such as ADAM (Kingma and Ba
[2014]) or L-BFGS (Liu and Nocedal [1989]). These optimization routines internally solve the objective function several times in the given space until it arrives at the set of parameters which minimizes the objective function. Moreover, the objective function involves a matrix inversion operation, where the size of the matrix depends upon total training points. Furthermore, solving regression problems using trained GPs also involves matrix inversion problem.  
The matrix inversion in general has a runtime complexity of $\mathcal{O}(N^3)$, therefore the training of GPs can be computationally challenging.

\begin{figure}[ht]
	\centering
	\includegraphics[ width=0.5\textwidth]{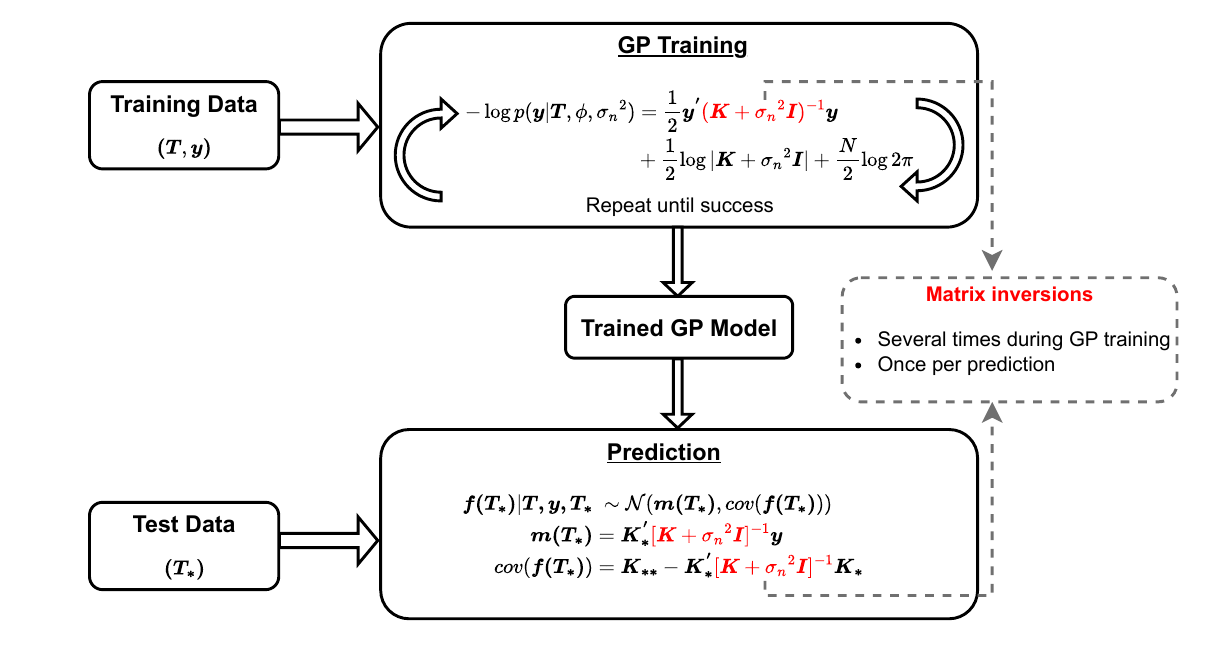}
	\caption[Caption used in list of tables]{A schematic diagram for GP training and prediction.}
	\label{fig:qgp_flowchart}
\end{figure}

\section{Quantum Gaussian Processes}
\label{sec:qgp}
%\section{Line Parameter Estimation - Quantum  Implementation}

In a standard QGP model we want to optimize the negative log marginal likelihood, which is defined in Eq. (\ref{MLE}), where
% \begin{align}
	% \begin{split}
		% NLML = \frac{1}{2} \mathbf{y}^{T}\left(\mathbf{K} + {\sigma_n}^2\mathbf{I}\right)^{-1}\mathbf{y}   
		%       + \frac{1}{2} \log |\mathbf{K} + {\sigma_n}^2\mathbf{I}| + \frac{N}{2}\log 2\pi.
		%     \label{NLML}
		% \end{split}
	% \end{align}
we compute the matrix inversion part,  $\frac{1}{2} \mathbf{y}^{T}(\mathbf{K} + {\sigma_n}^2\mathbf{I})^{-1}\mathbf{y}  $,  with HHL. During the optimization this part is computed several hundreds of times. In each time we get a speedup from the HHL over its classical counterpart and all the speedups cumulatively adds up over the whole optimization cycle.

\iffalse
A brief outline of QGP is presented in Fig. 
\begin{figure}[ht]
	\centering
	\includegraphics[width=0.5\textwidth]{QGP_Flow_chart.pdf}
	\caption[Caption used in list of tables]{A schematic diagram for QGP implementation.}
	\label{fig:qgp_flowchart}
\end{figure}
\fi

The main quantum algorithm used in QGP is HHL algorithm. In the following sections we describe a detail explanation of the HHL algorithm.

%\underline{\emph{Introduction}}

The given Linear system has to have a low condition number ${\displaystyle \kappa },$ and the Matrix $A$ must be $s$-sparse. This means $A$ must have at most $s$ non-zero entries per row or column. Solving an $s$-sparse system of size $N$ with a classical computer requires $\mathcal{ O }\left( N^3 s\kappa\log(1/\epsilon) \right)$ running time using the conjugate gradient method. Here, $\epsilon$ denotes the accuracy of the approximation. The computational complexity of HHL algorithm is $\mathcal{ O }\left(\log(N) s^2 \kappa^2 /\epsilon \right)$, therefore for a sparse matrix HHL provides up to exponential speedup over it's classical counterpart.

It is assumed that the user of the algorithm is interested in the result of a scalar measurement on the solution vector, instead of the values of the solution vector itself. So, it assumes that the user is not interested in the values of ${\displaystyle {\overrightarrow {x}}}$ itself, but rather the result of applying some operator ${\displaystyle M}$ onto x, ${\displaystyle \langle x|M|x\rangle }$. Hence, while the classical algorithm returns the full solution, the HHL can only approximate functions of the solution vector.

Also, matrix ${\displaystyle A}$ should be Hermitian so that it can be converted into a unitary operator, under the assumptions of efficient oracles for loading the data, Hamiltonian simulation and computing a function of the solution. 

\subsection{Mathematical preliminaries}

The first step towards solving a system of linear equations is to encode the problem in the quantum computer. By rescaling the system, we can assume $\vec{b}$ and $\vec{x}$ to be normalised and map them to the respective quantum states $|b\rangle$ and $|x\rangle$. Usually the mapping used is such that $i^{th}$ component of $\vec{b}$ (resp. $\vec{x}$) corresponds to the amplitude of the $i^{th}$ basis state of the quantum state $|b\rangle$ (resp. $|x\rangle$). From now on, we will focus on the rescaled problem
\begin{equation}
	A|x\rangle=|b\rangle .
\end{equation}

Since $A$ is Hermitian, it has a spectral decomposition
\begin{equation}
	A=\sum_{j=0}^{N-1}\lambda_{j}|u_{j}\rangle\langle u_{j}|,\quad \lambda_{j}\in\mathbb{ R },   
\end{equation}
where $|u_{j}\rangle$ is the $j^{th}$ eigenvector of $A$ with respective eigenvalue $\lambda_{j}$.  It can be written as a the sum of the outer products of its eigenvectors, scaled by its eigenvalues. Therefore, we can write the inverse of A as
\begin{equation}
	A^{-1}=\sum_{j=0}^{N-1}\lambda_{j}^{-1}|u_{j}\rangle\langle u_{j}| .    
\end{equation}

Since $A$ is invertible and Hermitian, it must have an orthogonal basis of eigenvectors, and thus we can write $b$ in the eigenbasis of $A$ as
\begin{equation}
	|b\rangle=\sum_{j=0}^{N-1}b_{j}|u_{j}\rangle,\quad b_{j}\in\mathbb{ C } .
\end{equation}
The goal of the HHL is to exit the algorithm with the readout register in the state
\begin{equation}
	|x\rangle=A^{-1}|b\rangle=\sum_{j=0}^{N-1}\lambda_{j}^{-1}b_{j}|u_{j}\rangle .  
\end{equation}

%Note that here we already have an implicit normalisation constant since we are talking about a quantum state.

\subsection{The HHL algorithm}\label{subsec:HHL_algorithm}

The algorithm uses three quantum registers, all of them set to $|0\rangle $ at the beginning of the algorithm. The first register, which is denoted with the subindex $n_{l}$, is used to store a binary representation of the eigenvalues of $A$. A second register, denoted by $n_{b}$, contains the vector solution with $N=2^{n_{b}}$. There is an extra register for the auxiliary qubits. These are qubits used as intermediate steps in the individual computations but will be ignored in the following description since they are set to $|0\rangle $ at the beginning of each computation and restored back to the $|0\rangle $ state at the end of the individual operation.

\begin{figure}[ht]
	\centering
	\includegraphics[width=0.5\textwidth]{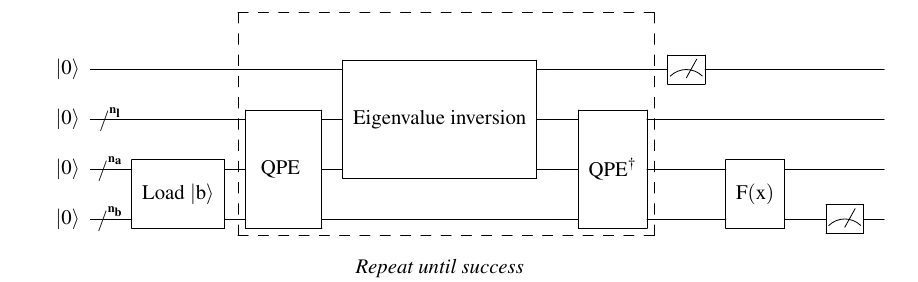}
	\caption[Caption used in list of tables]{A schematic diagram for HHL algorithm.}
	\label{fig:hhl_circuit}
\end{figure}

In Fig.\ref{fig:hhl_circuit},  a schematic diagram for HHL algorithm is presented.

% \begin{figure}[h]
	%     \begin{center}
		%        \resizebox{\textwidth}{!}{
			%             \input{latex_files/HHL.tex}
			%         }
		%     \parbox{0.9\textwidth}{
			%     \caption[Caption used in list of tables]{A schematic diagram for HHL algorithm.}\label{fig:hhl_circuit}}
		%     \end{center}
	% \end{figure}

1.  Load the data $|b\rangle\in\mathbb{ C }^{N}$. That is, perform the transformation
\begin{equation}
	|0\rangle _{n_{b}} \mapsto |b\rangle _{n_{b}} . 
\end{equation}
2. Apply Quantum Phase Estimation (QPE) with
\begin{equation}
	U = e ^ { i A t } := \sum _{j=0}^{N-1}e ^ { i \lambda _ { j } t } |u_{j}\rangle\langle u_{j}|.   
\end{equation}	

The quantum state of the register expressed in the eigenbasis of $A$ is now
\begin{equation}
	\sum_{j=0}^{N-1} b _ { j } |\lambda _ {j }\rangle_{n_{l}} |u_{j}\rangle_{n_{b}}, 
\end{equation}
where $|\lambda _ {j }\rangle_{n_{l}}$ is the $n_{l}$-bit binary representation of $\lambda _ {j }$.

3. Add an auxiliary qubit and apply a rotation conditioned on $|\lambda_{ j }\rangle$,
\begin{equation}
	\sum_{j=0}^{N-1} b _ { j } |\lambda _ { j }\rangle_{n_{l}}|u_{j}\rangle_{n_{b}} \left( \sqrt { 1 - \frac { C^{2}  } { \lambda _ { j } ^ { 2 } } } |0\rangle + \frac { C } { \lambda _ { j } } |1\rangle \right),  
\end{equation}
where $C$ is a normalisation constant, and, as expressed in the current form above, should be less than the smallest eigenvalue $\lambda_{min}$ in magnitude, i.e., $|C| < \lambda_{min}$.

4. Apply QPE$^{\dagger}$. Ignoring possible errors from QPE, this results in
\begin{equation}
	\sum_{j=0}^{N-1} b _ { j } |0\rangle_{n_{l}}|u_{j}\rangle_{n_{b}} \left( \sqrt { 1 - \frac {C^{2}  } { \lambda _ { j } ^ { 2 } } } |0\rangle + \frac { C } { \lambda _ { j } } |1\rangle \right) .
\end{equation}

5. Measure the auxiliary qubit in the computational basis. If the outcome is $1$, the register is in the post-measurement state
\begin{equation}
	\left( \sqrt { \frac { 1 } { \sum_{j=0}^{N-1} \left| b _ { j } \right| ^ { 2 } / \left| \lambda _ { j } \right| ^ { 2 } } } \right) \sum _{j=0}^{N-1} \frac{b _ { j }}{\lambda _ { j }} |0\rangle_{n_{l}}|u_{j}\rangle_{n_{b}},  
\end{equation}
which up to a normalisation factor corresponds to the solution.

6. Apply an observable $M$ to calculate 
\begin{equation}
	F(x):=\langle x|M|x\rangle .
\end{equation}

\subsection{Quantum phase estimation (QPE) within HHL} 
Quantum phase estimation is the core quantum subroutine in the HHL algorithm. Roughly speaking for given a unitary $U$ with eigenvector $|\psi\rangle_{m}$ and eigenvalue $e^{2\pi i\theta}$, QPE finds $\theta$. 

More generally let $U\in\mathbb{ C }^{2^{m}\times 2^{m}}$ be a unitary matrix and let $|\psi\rangle_{m}\in\mathbb{ C }^{2^{m}}$ be one of its eigenvectors with respective eigenvalue $e^{2\pi i\theta}$. The QPE algorithm takes as inputs the unitary gate for $U$ and the state $|0\rangle_{n}|\psi\rangle_{m}$ and returns the state $|\tilde{\theta}\rangle_{n}|\psi\rangle_{m}$. Here $\tilde{\theta}$ denotes a binary approximation to $2^{n}\theta$ and the $n$ subscript denotes it has been truncated to $n$ digits.
\begin{equation}
	\operatorname { QPE } \left( U , |0\rangle_{n}|\psi\rangle_{m} \right) = |\tilde{\theta}\rangle_{n}|\psi\rangle_{m}.
\end{equation}

For the HHL we use QPE with $U = e ^ { i A t }$, where $A$ is the matrix associated to the system we want to solve. In this case, 
\begin{equation}
	e ^ { i A t } = \sum_{j=0}^{N-1}e^{i\lambda_{j}t}|u_{j}\rangle\langle u_{j}|.
\end{equation}
Then, for the eigenvector $|u_{j}\rangle_{n_{b}}$, which has eigenvalue $e ^ { i \lambda _ { j } t }$, QPE will output $|\tilde{\lambda }_ { j }\rangle_{n_{l}}|u_{j}\rangle_{n_{b}}$. Where $\tilde{\lambda }_ { j }$ represents an $n_{l}$-bit binary approximation to $2^{n_l}\frac{\lambda_ { j }t}{2\pi}$. Therefore, if each $\lambda_{j}$ can be exactly represented with $n_{l}$ bits,
\begin{equation}
	\operatorname { QPE } \left( e ^ { i A t } , \sum_{j=0}^{N-1}b_{j}|0\rangle_{n_{l}}|u_{j}\rangle_{n_{b}} \right) = \sum_{j=0}^{N-1}b_{j}|\lambda_{j}\rangle_{n_{l}}|u_{j}\rangle_{n_{b}}.    
\end{equation}

\section{Application to Line Parameter Estimation in Electrical Networks}
\label{sec:gp-lpe}

\subsection{Problem Formulation}
The knowledge on electrical line parameters may be outdated, incorrectly recorded or even unknown, especially for medium to low voltage grids. With the availability of measurements, it was demonstrated in \cite{gp-lpe} that these line parameters can be estimated in a physics-informed data-driven manner. In this section, we briefly discuss this approach of GP based network line parameter estimation (LPE) and measurement signal predictions. Interested readers may refer to  \cite{gp-lpe} for details on the approach itself.

\begin{figure}[!ht]
	\centering
	\includegraphics[width=0.4\textwidth]{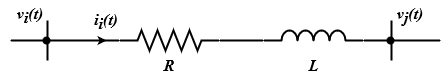}
	\caption[Caption used in list of tables]{Single line diagram of a two node network.}
	\label{pi_rlc}
\end{figure}

The commonly used equivalent circuit representation of a short (up to 80km) to medium length (80-250km) transmission line in power system is given by a two-port Pi-model whose parameters correspond to the positive sequence equivalent circuit of transmission line consisting of resistance $R$, series inductance $L$ and charging capacitance \cite{Abur_Exposito}. For short lines, with line lengths less than 80km, the shunt charging capacitance are often ignored, therefore the pi-model reduces to equivalent circuit  as shown in Fig. \ref{pi_rlc}, where $v_i(t)$ is the sending end voltage and $v_j(t)$ is the receiving end voltage, $i_i(t)$ is the branch current. With this assumption, the dynamics of the line is governed by:

\begin{align}
	v_i(t)  = Ri_{i}(t) + L\frac{di_{i}(t)}{dt} + v_j(t) .
	\label{short_line}
\end{align}

Assuming the measurement samples of $v_i(t), v_j(t)$ and $i_i(t)$ are available, and that the measurement noise are random errors following independent identical Gaussian distribution (i.i.d), with variances of measurement noise signals  ${\sigma_{v_i}}^2$, ${\sigma_{v_j}}^2$, ${\sigma_{i_i}}^2$ respectively. For estimating the line parameters, we model each measurement signal as a GP over time, as shown in \ref{meas_gp}. 

\begin{equation}
	\begin{aligned}
		\mathbf{v_i} &~ \sim\mathcal{GP}(0, k_{v_iv_i}(\mathbf{t_{v_i}},\mathbf{t_{v_i}}) + {\sigma_{v_i}}^2\mathbf{I}) ,\\
		\mathbf{v_j} &~ \sim\mathcal{GP}(0, k_{v_jv_j}(\mathbf{t_{v_j}},\mathbf{t_{v_j}}) + {\sigma_{v_j}}^2\mathbf{I}) ,\\
		\mathbf{i_i} &~ \sim\mathcal{GP}(0, k_{i_ii_i}(\mathbf{t_{i_i}},\mathbf{t_{i_i}}) + {\sigma_{i_i}}^2\mathbf{I}) .
		\label{meas_gp}
	\end{aligned}
\end{equation}
The measurements with time-stamp \{$\mathbf{y_{v_i}, t_{v_i}}$\}, \{$\mathbf{y_{v_j}, t_{v_j}}$\} and \{$\mathbf{y_{i_i}, t_{i_i}}$\} are used as training data, where $\mathbf{y_{v_i}, y_{v_j}, y_{i_i}}$ are vector of measurements observed at vector of time instants $\mathbf{t_{v_i}, t_{v_j}, t_{i_i}}$ respectively. The vectors $\mathbf{t_{v_i}, t_{v_j}, t_{i_i}}$ can be different from one another due to non-uniform sampling rates of measuring devices. 

Starting with base RBF kernels for $\mathbf{i_i}$ and  $\mathbf{v_j}$, the other covariances can be derived from the correlation arising due to physical laws between $v_i(t)$, $i_i(t)$ and $v_j(t)$, which is given by \ref{short_line}. By doing this, the line parameters become part of the kernels and therefore can be learnt together with other GP hyper-parameters during training. This procedure is derived and explained in \cite{gp-lpe}, helps in simultaneous learning of multiple output in a single formulation given by \eqref{joint_mean} and \eqref{joint_cov}.

%Using the measurements of $v_i, v_j$ and $i_i$ as training data, each of these signal can be modeled as a GP. 
%Previously in \cite{gp-lpe} it was shown that given the physics-based equation describing relation between electrical signals ($v_i, v_j, i_i$) and parameters ($R, L$) of a line, when these electrical signals are represented as GPs, the measurements of these quantities can be used as training data to compute the line parameters. 
%This approach is learning of physics-informed linear equations, that is, equations governed by Kirchoff's voltage and current laws, and thereby to learn the parameters of these equations. We use the time domain dynamic equation of a line which follows the laws of conservation of energy and electric charge, to learn directly the line parameters. 

Once the hyper-parameters are optimized, the learned GP models of the respective measurement signals can be used for predictions at test data-points say $\mathbf{t_*}$ using following equations 
\begin{align}
	m_{s}(\mathbf{t_*}) &= q_{s}^{'} [\mathbf{K} + {\sigma_n}^2 \mathbf{I}]^{-1}\mathbf{y} \label{joint_mean} ,\\ 
	cov_{s}(\mathbf{t_*}) &= \mathbf{k_{s}(\mathbf{t_*}, \mathbf{t_*}) - q_{s}^{'} [\mathbf{K} + {\sigma_n}^2 \mathbf{I}]^{-1}q_{s}} , \label{joint_cov}
\end{align}    
where $s = i_i, v_i, v_j$, and 
\begin{align*}    
	q_{i_i}^{'} = [\mathbf{k_{{i_i}{i_i}}}(\mathbf{t_*}, \mathbf{t_{i_i}}) \;\;\; \mathbf{k_{{i_i}{v_j}}}  (\mathbf{t_*}, \mathbf{t_{v_j}}) \;\;\; \mathbf{k_{{i_i}{v_i}}}(\mathbf{t_*}, \mathbf{t_{v_i}}) ] , \\
	q_{v_j}^{'} = [\mathbf{k_{{v_j}{i_i}}}(\mathbf{t_*}, \mathbf{t_{i_i}}) \;\;\; \mathbf{k_{{v_j}{v_j}}}  (\mathbf{t_*}, \mathbf{t_{v_j}}) \;\;\; \mathbf{k_{{v_j}{v_i}}}(\mathbf{t_*}, \mathbf{t_{v_i}}) ] ,
	\\
	q_{v_i}^{'} = [\mathbf{k_{{v_i}{i_i}}}(\mathbf{t_*}, \mathbf{t_{i_i}}) \;\;\; \mathbf{k_{{v_i}{v_j}}}  (\mathbf{t_*}, \mathbf{t_{v_j}}) \;\;\; \mathbf{k_{{v_i}{v_i}}}(\mathbf{t_*}, \mathbf{t_{v_i}}) ] .
\end{align*}

\subsection{Test set-up}
\begin{figure}[!ht]
	\centering
	\includegraphics[width=0.48\textwidth]{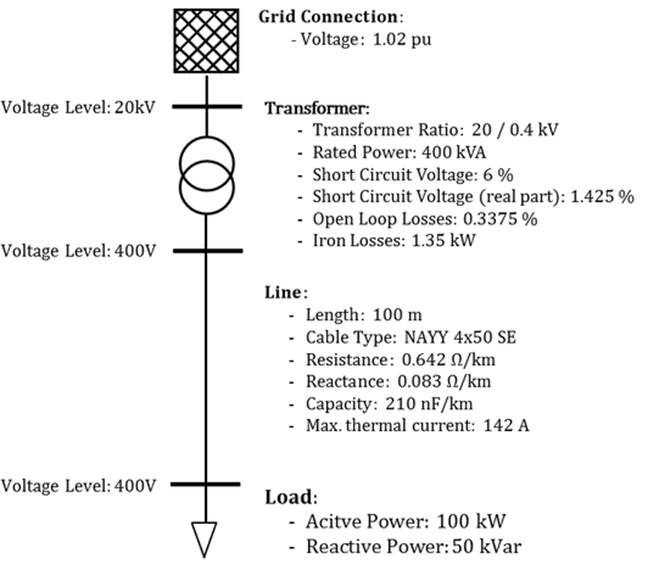}
	\caption[Caption used in list of tables]{The test network  \cite{pandapower.2018}}
	\label{network}
\end{figure}

\begin{figure}[ht]
	\centering
	\includegraphics[height=0.7\textwidth]{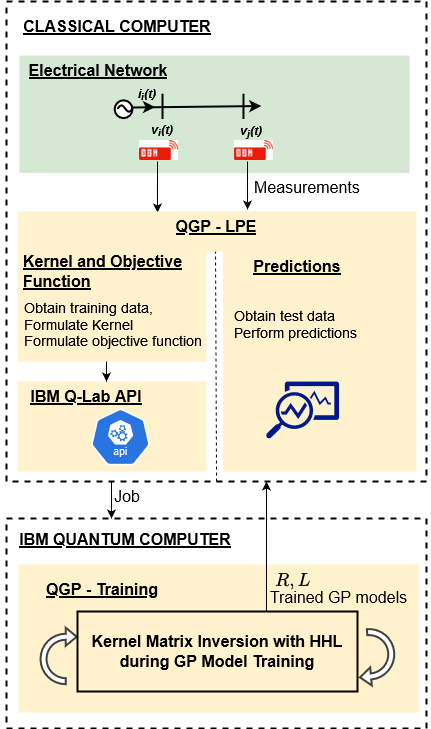}
	\caption[Caption used in list of tables]{Proof-of-concept for quantum Gaussian process based line parameter estimation (QGP-LPE).}
	\label{fig:qgp_exp_setup}
\end{figure}

The test set-up used for multi-output quantum Gaussian process based line parameter estimation (QGP-LPE) is pictorially depicted in Fig. \ref{fig:qgp_exp_setup}. It consists of a classical computer which is a Lenovo ThinkPad with intel core i5, 8th Gen processor and IBM Quantum lab API for accessing Qsam simulator as well as Quantum computer hardware hosted by IBM. For this test, we consider a short-distance line of a simple test network as shown in the Fig. \ref{network} and implement the line parameter estimation using classical and quantum variants of GP for a single phase. We assume a balanced network and steady-state operating conditions at nominal network frequency. The line end voltages and branch currents of the test network are simulated at steady state on a classical computer. To emulate noisy measurement signals, random Gaussian errors are added to the computed signals. Samples from these corrupted signals serve as training data for the GP based line parameter estimation as explained in \ref{sec:gp-lpe}. For QGP, the kernel matrix encoding the training data as well as the objective function for hyper-parameter optimization is formulated on classical computer and a job session is evoked using IBM's quantum lab API. In this job, the optimization of hyper-parameter and consequently estimation of $R$ and $L$ values take place on quantum computer, after which these results are sent back to classical computer. The optimized hyper-parameters of GPs along with $R$ and $L$ can eventually be used to perform measurement signal predictions on classical computer. 

%The network has the following line parameters, namely, resistance (R)= 0.064 ohm and inductance (L)=  $2.6 \times 10^{-5}$. 
We consider 11 samples each for load voltage $v_j$ and branch current $i_i$, and 10 samples for sending end voltage $v_i$, therefore we need a total of 32 training data points for line parameter estimation, and a size of $32 \times 32$ Kernel matrix. 

%\subsection{Challenges}
% I moved this paragraph under next sub-section

		\subsection{QGP Implementation}\label{subsec:qgp_implementation}
		Although quantum computing has shown potential for a significant speedups over its classical counterparts, but current quantum resources is limited. For e.g., HHL requires a large gate depth which is only possible to implement with a fault tolerant quantum computer which is still not available currently. 
		During optimization of NLML in QGP, we encounter a very high condition number of the matrix $\left(\mathbf{K} + {\sigma_n}^2\mathbf{I}\right)^{-1}$, usually $\mathcal{O}(2^{30})$ considering instantaneous measurement values from a standard 50 Hz two node network. The high condition number gives rise to a large circuit width and also enormously large circuit depth in HHL circuit; therefore it becomes almost impossible to implement HHL with such high condition number.
		
		Therefore we reformulate the problem in a way such that it is compatible to HHL and can be possibly implemented in a quantum computer with limited resources.  We adopt the following steps for implementing a quantum inspired Gaussian Process for the line parameter estimation.
		
		\textbf{Step 1 : applying HHL for computing the first term of the log marginal likelihood function}
		
		In Eq. (\ref{MLE}), the term $ \mathbf{y}^{T}(\mathbf{K} + {\sigma_n}^2\mathbf{I})^{-1}\mathbf{y}$, can be written as:
		$\langle \mathbf{y} |\mathcal{K}| \mathbf{y}\rangle$, where $\mathcal{K}= (\mathbf{K} + {\sigma_n}^2\mathbf{I})^-1$, which a Hermitian and positive semi-definite (PSD) matrix. For a given matrix $A$ and vector $|b\rangle$, from HHL we can directly compute the norm $||x||^2$, where the vector $|x\rangle=A^{-1}|b\rangle$. For our case we first break the PSD matrix $\mathcal{K}$ into two Hermitian matrices $\kappa$ and $\kappa^\dagger$ with Schutz decomposition. Then for a given $\kappa$  and $\mathbf{y}$ we compute the norm $||z||^2= \langle \mathbf{y} |\mathcal{K}| \mathbf{y}\rangle$ where $z=\kappa^{-1}|\mathbf{y}\rangle$ with HHL.
		
		\textbf{Step 2 : conditioning of the kernel matrix to reducing the condition number}
		
		As mention above the condition numbers of $\kappa$ is very high during optimization, therefore its very hard to implement HHL with such a high condition number. Therefore to tackle this issue we use few mathematical  tricks. Firstly we implement a matrix conditioning method to bring the condition number down. We add a strong regularization factor with the kernel matrix  $\mathbf{K}$ defined above. Then  we adopt a matrix conditioning scheme described by  Braatz and Morari \cite{braatz1994minimizing} where  we scale $\kappa$, a Positive semidefinite matrix, to $\kappa^\prime=(D \kappa D)$ with a diagonal matrix $D= \text{diag} \left( \kappa_{11}^{-1/2}, \kappa_{22}^{-1/2}, ... , \kappa_{nn}^{-1/2}\right)$, which brings down the condition number of $\kappa$ matrix $\mathcal{O}(2^{9})$ and also the resultant matrix is also positive semi-definite. Due to relatively low condition number it is implementable in the current quantum computers. After computing the inner product, $\langle \mathbf{y} |\mathcal{K}^\prime| \mathbf{y}\rangle$ with scaled matrix $\kappa^\prime$ we rescale back the inner product by multiplying a scale factor $d_{min}$, where $d_{min}$ is the minimum value of the diagonal matrix $D^{-1}$. 
		
		\textbf{Step 3 : applying AQC for gate-depth reduction}
		
		Even after applying step 2, the typical gate depth of HHL is order of billions, which is still beyond the reach of current quantum hardware or simulators. The quantum phase estimation (QPE) part in HHL produces the enormous depth. Therefore we implement a special technique for reducing the gate depth called Approximate Quantum Compiling (AQC), which is described in Appendix \ref{appendix:AQC}. After applying the AQC in the QPE part of HHL we significantly reduced the circuit depth (from order of billions to order of few hundreds).
		
		Although AQC compiling is a very efficient algorithm to reduce gate depth but it fails to converge for high-dimensional $U\in SU(2^n)$ matrices due to Barren Plateau phenomena. Due to overcome this limitation we put an upper limit ($\mathcal{E}$) in the number of evaluation qubits in the QPE.  If the condition number of the corresponding matrix involved in HHL crosses a certain limit and therefore the number of evaluation qubits in QPE exceeds the upper limit ($\mathcal{E}$), we evaluate the QPE using only $\mathcal{E}$ qubits. For our particular experiment we keep $\mathcal{E} =8$, therefore QPE will compute the Eigen values of the corresponding matrix up to $8$-binary places.
		
		Note that, in our case we only are interested in the value of the norm $||x||^2$, where $|x\rangle=K^{-1}|y\rangle=\sum_{j=0}^{N-1}\lambda_{j}^{-1}y_{j}|u_{j}\rangle $. After the step 3 (eigen value inversion) of the HHL algorithm (see subsection \ref{subsec:HHL_algorithm}), the auxiliary qubit is in the state $\sum_{j=0}^{N-1} \left( \sqrt { 1 - \frac { C^{2}  } { \lambda _ { j } ^ { 2 } } } |0\rangle + \frac { C } { \lambda _ { j } } |1\rangle \right).$ Therefore, we measure the probability of getting the auxiliary qubit in state $|1\rangle$, which is the desired norm $||x||^2$ with a normalization constant $C$.

%\section{ Results}
\section{Result summary and discussions}
\label{sec:results_diss}
The line parameter estimation on test network explained in previous sections are performed using QGP on qasm simulator and on a real hardware (IBM Auckland). The same line parameter estimation is carried out on classical computer with classical GP and the results are summarized in Table \ref{table:quantum_result}. 
% \begin{table*}[ht]
	% \begin{center}
		% \begin{tabular}{c c c c c c} 
			%  \toprule
			%  \# samples &  Parameter &  Actual &  Qasm &  IBM Auckland &  Classical GP \\
			%  &  &  &   &  &  
			%  \\ [0.5ex] 
			%  \midrule
			% $n_{v_i}$ = 10,  $n_{i_i}$ =11 &  $R (\Omega)$ &  0.064 &            0.042 &               0.089 &                      0.064 \\
			%  $n_{v_j}$ = 11 &  $L (H)$ &  $2.64 \times 10^{-5}$  &     $6.07 \times 10^{-5}$    &             $5.63 \times 10^{-5}$   &                      $2.63 \times 10^{-5}$ \\ [1ex] 
			%  \bottomrule
			% \end{tabular}
		% \caption{\label{table:quantum_result} The line parameter estimation result is summarised. We compare the results obtained from Qasm simulator and real hardware IBM Auckland with the results obtained from classical GP.}
		% \end{center}
	% \end{table*}

\begin{table*}[ht]
	\begin{center}
		\begin{tabular}{c c c c c c c} 
			\toprule
			\# samples &  Parameter &  Actual &  Classical GP & Qasm &  IBM Auckland &  Absolute Error \\
			&  &  &   &  & & with QGP (IBM Auckland) 
			\\ [0.5ex] 
			\midrule
			$n_{v_i}$ = 10,  $n_{i_i}$ =11 &  $R (\Omega)$ &  0.064 &   0.064 &    0.042 &               0.089      &     39.06\%        \\
			$n_{v_j}$ = 11 &  $L (H)$ &  $2.64 \times 10^{-5}$  &  $2.63 \times 10^{-5}$ &   $6.07 \times 10^{-5}$    &      $5.63 \times 10^{-5}$   & 113.28\% \\ [1ex] 
			\bottomrule
		\end{tabular}
		\caption{\label{table:quantum_result} The line parameter estimation result is summarised. We compare the learnt $R$ and $L$ results obtained from Qasm simulator and real hardware IBM Auckland with the results obtained from classical GP.}
	\end{center}
\end{table*}

The classical GP on classical computer gives the closest estimation of $R$ and $L$ parameters. The Qsam as well as the real quantum hardware produce results which are not highly accurate, however they are on the same decimal range as of actual values of $R$ and $L$ respectively. Note that, due to the limitation of quantum hardware resources and time constraint, we use maximum 100 iterations during the QGP optimization process for the real hardware experiments. In actual cases the complete optimization process can involve hundreds of iterations, therefore the result is not very accurate. Also, we reduce the circuit depth to order of hundreds from actual depth of billions so this heavy approximation also affects the result accuracy. 

\begin{figure}[ht]
	\centering
	\includegraphics[width=0.48\textwidth]{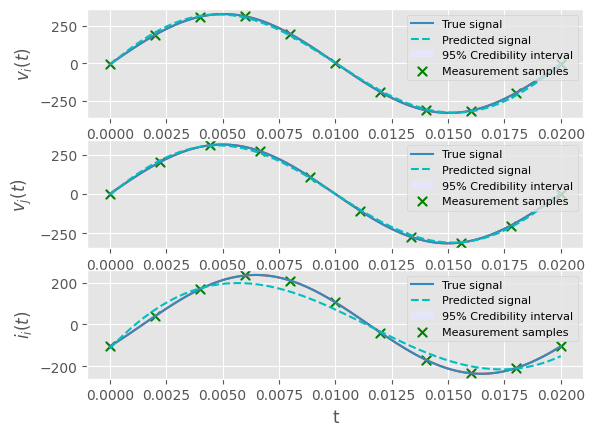}
	\caption[Caption used in list of tables]{A pictorial representation of the QGP regression of measurement signals using hyper-parameters computed by IBM Auckland (dashed lines) in comparison to the true signal solid lines.}
	\label{QGP_simulation}
\end{figure}

Further, with the optimized hyper-parameters, the GP models of measured signals can be used for regression which may provide useful inputs other applications, such as state-estimation. We test this regression by imputing the measurement signals over 200 points over a signal cycle. Fig \ref{QGP_simulation} gives a graphical representation of our QGP regression results with IBM Auckland. It displays that the other hyper-parameters of GPs are also optimized well enough for ensuring close imputations.

From this study, following points are worth highlighting:
\begin{itemize}
	\item The results obtained from the noise-less simulator (Qsam) and the noisy intermediate quantum computer (IBM Auckland) are very close-by, displaying the robustness of the approach to the quantum hardware noise. 
	\item The QGP in this experiment inverted a matrix of size 32x32, which is a fairly large size of matrix on which HHL is tested on NISQ device. 
	\item The QGP optimized a total of 9 hyper-parameters including $R$, $L$, and other hyper-parameters of $v_i$, $i_i$ and $v_j$ GP models. The prediction plots of GPs in Fig. \ref{QGP_simulation} displays that the GP models thus constructed with optimized hyper-parameters are fairly precise, leading to relatively accurate predictions.
\end{itemize}

The proposed QGP approach showcases promising outcome for the line parameter application. 

%\underline{Feasibility Check:} \\
Although this approach theoretically promises speed-up, the actual execution time may be far from being suitable for real-world application. Moreover, mitigation of errors in estimation and large gate depth due to high condition number matrix must be further improved for making it further suitable for achieving quantum utility for critical grid applications. The accessibility of quantum computers, cost of executing a job on quantum computers, as well as lack of professionals in this interdisciplinary field are other challenges to be addressed in future. For the immediate future, one may foresee quantum computation to aid classical computers for offline applications.

\section{Conclusions}
\label{sec:conclusion}
Gaussian process is powerful tool in modeling complex data in machine learning, albeit the computational burden associated with the technique hinders it's application. This article proposes a multi-output Quantum Gaussian process, where the kernel matrix inversion of GP is replaced by the HHL algorithm, which is known for providing significant theoretical speed-up. However, the implementation of HHL on NISQ devices is challanging due to its high circuit depth. In this article, the high circuit depth requirement of HHL (therefore QGP) is tackled by applying approximate quantum compiling. We further curate the developed QGP for electrical network line parameter estimation and measurement signals prediction, which is demonstrated on noise-less quantum-simulator and real quantum hardware (IBM Auckland), thus showcasing a proof-of-concept for a real-world machine learning problem using multi-output QGP on NISQ device. It is observed that for the test network under study, a total of 32 samples are needed for learning the measurements, thus the Kernel is of size 32x32. Moreover, depending upon the condition number of the Kernel matrix, up to 13-qubit  HHL-circuit has been implemented in real quantum hardware, which is significantly larger in size compared to the existing works \cite{qc_pf_real_test} in the similar field. 

\bibliographystyle{apsrev4-1}
\bibliography{ref}

\appendix{
\section{Approximate Quantum Compiling}
\label{appendix:AQC}
	During computation of the line parameter using HHL, we encounter quantum circuits with great depth which is impossible to implement in the current quantum computers. In quantum circuits each individual quantum gate can be represented as small unitary matrices. Therefore the whole quantum circuit can be thought as a giant unitary matrix constructed by multiplying billions of such small unitary matrices. The general problem of gate compiling is to find an exact combination of universal gates, one and two-qubit  (CNOT) gates, such that it exactly produces a given arbitrary matrix $U\in SU(2^n)$, where $n$ is the number of qubits. The minimum number of CNOT gates needed to compile a general gate is equal to $\lceil \tfrac{1}{4}(4^n-3n-1)\rceil$ \cite{shende2004minimal}, which grows exponentially with the circuit width.  \\
	To overcome the above problem we are interested in an approximate representation of an exact unitary matrix such that we can move beyond the lower bounds of the above CNOT cost. without over-approximating the original circuit. In other words, we want to find an approximate circuit that satisfies the hardware constraint, i.e., reduced number of CNOT gates, and is a close  approximation of the exact target circuit.  Let $U\in SU(2^n)$ be the unitary matrix induced by the ordered gate sequence of a quantum circuit $\mathcal{S}$ and let $V\in SU(2^n)$ be another unitary matrix, associated to the approximate circuit. Then the pertinent metric is defined as the distance induced by the Frobenius norm, $|| V -U ||_F$. 
	Ref.~\cite{AQC} provides a method for finding such approximate circuits with a specified number of CNOT gates, called Approximate Quantum Compiling (AQC) \cite{AQC}. This AQC technique has been recently applied to solve a real life energy portfolio risk analysis problem in article \cite{ghosh2024energy}.

\section{Hardware specification of the IBM device used for QGP}\label{device specifications}

A topology graph of IBM Auckland is provided in Fig.~\ref{fig:mumbai}, while Table \ref{table:ibm_mumbai} contains some of its specifications. 

\begin{figure}[h!]
	\centering
	\includegraphics[width=0.8\linewidth]{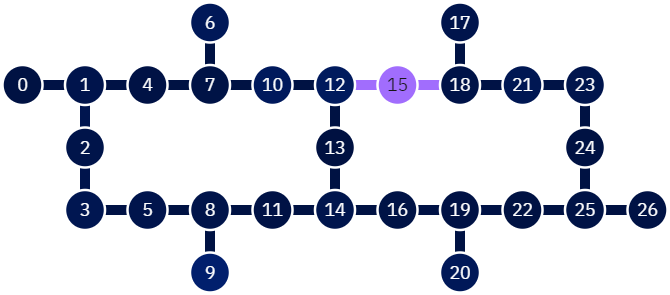}
	\caption{The qubit topology graph of IBM Auckland, which is 27-qubit quantum device. Lighter colors on qubits represent higher readout assignment errors, while on edges represent higher CNOT errors.}
	\label{fig:mumbai}
\end{figure}

\begin{table}[ht]
	\begin{center}
		\begin{tabular}{c c c c c c} 
			\toprule
			Qubit &  T1 &  T2 &  Freq &  Anharm &  Readout \\
			&  (us) &  (us) &  (GHz) & (GHz) &  assignment error
			\\ [0.5ex] 
			\midrule
			0 &   98.064 & 174.081 &            4.933 &               -0.343 &                      0.006 \\
			1 &  112.497 &  166.727 &            5.074 &               -0.342 &                      0.011 \\
			2 &   81.551 &  220.360 &            5.006 &               -0.343 &                      0.010 \\
			3 &  207.860 & 120.070 &            4.897 &               -0.346 &                      0.014 \\
			4 &  236.635 & 247.988 &            4.920 &               -0.345 &                      0.007 \\
			5 &  149.642 &  62.488 &            4.993 &               -0.344 &                      0.011 \\
			6 &  160.692 & 177.788 &            5.007 &               -0.344 &                      0.018 \\
			7 &  110.387 & 176.673 &            4.828 &               -0.346 &                      0.008 \\
			8 &   93.938 &  78.614 &            5.204 &               -0.341 &                      0.010 \\
			9 &   88.851 &  211.071 &            5.088 &               -0.343 &                      0.032 \\
			10 &  154.307 &  48.902 &            4.727 &               -0.351 &                      0.020 \\
			11 &  148.840 & 116.853 &            5.055 &               -0.342 &                      0.007 \\
			12 &  166.461 &  180.080 &            4.890 &               -0.345 &                      0.022 \\
			13 &  200.983 & 27.544 &            5.017 &               -0.344 &                      0.005 \\
			14 &  130.394 & 209.241 &            5.167 &               -0.342 &                      0.013 \\
			15 &   98.655 &         undefined &            4.988 &               -0.343 &                      0.218 \\
			16 &  178.984 & 142.529 &            4.970 &               -0.344 &                      0.008 \\
			17 &  118.023 &  134.118 &            5.025 &               -0.343 &                      0.012 \\
			18 &  111.703 & 179.733 &            4.788 &               -0.349 &                      0.009 \\
			19 &  118.877 & 20.957 &            4.806 &               -0.347 &                      0.007 \\
			20 &  138.494 & 162.749 &            4.692 &               -0.349 &                      0.019 \\
			21 &   43.769 &   67.989 &            5.046 &               -0.342 &                      0.015 \\
			22 &  141.627 & 40.477 &            4.968 &               -0.344 &                      0.006 \\
			23 &  110.477 & 38.698 &            4.868 &               -0.345 &                      0.012 \\
			24 &  132.320 & 17.846 &            4.956 &               -0.343 &                      0.006 \\
			25 &  150.855 &  230.002 &            5.077 &               -0.343 &                      0.008 \\
			26 &  155.680 & 17.676 &            4.856 &               -0.345 &                      0.007 \\ [1ex] 
			\bottomrule
		\end{tabular}
		\caption{\label{table:ibm_mumbai} Some of the hardware specifications and calibration details of the {IBM Auckland} device are collected in the above table.}
	\end{center}
\end{table}
}

\end{document}